# Crystal structure and the magnetic properties of the 5$d$ transition metal oxide $A$OsO$_4$ ($A$ = K, Rb, Cs)

Jun-ichi Yamaura[1] and Zenji Hiroi[2]

[1]Materials Research Center for Element Strategy, Tokyo Institute of Technology, Yokohama, Kanagawa 226-8503, Japan
[2]Institute for Solid State Physics, The University of Tokyo, Kashiwa, Chiba 277-8581

We synthesized the 5$d^1$-transition metal oxides $A$OsO$_4$ ($A$ = K, Rb, Cs) by solid-state reaction, and performed structure determination and magnetic and heat capacity measurements. It was found that they crystallize in a scheelite ($A$ = K and Rb) or a quasi-scheelite structure ($A$ = Cs) comprising of distorted diamond lattices of septivalent Os ($d^1$) ions tetrahedrally coordinated by four oxide ions without local inversion symmetry; hence an antisymmetric spin–orbit coupling is expected in the crystals. The K and Rb compounds have Weiss temperatures of $\theta$ = −66 and −18 K, effective magnetic moments of $\mu_{\text{eff}}$ = 1.44 and 1.45 $\mu_B$/Os, and antiferromagnetic transition temperatures of $T_N$ = 36.9 and 21.0 K, respectively. In contrast, the Cs compound has $\theta$ = 12 K and $\mu_{\text{eff}}$ = 0.8 $\mu_B$/Os without magnetic transition above 2 K, instead exhibiting a first-order structural transition at $T_s$ = 152.5 K. The decline of the Os moment from 1.73 $\mu_B$/Os for the simple $d^1$ spin, particularly for Cs, is likely to originate from the antiparallel orbital moment, although the spin–orbit coupling is generally quenched in the low-lying $e$ orbitals.

# I. INTRODUCTION

Transition metal oxides (TMOs) have been long used for their rich variety of electrical and magnetic properties arising from the interplay among spin, orbital, and charge degrees of freedom. Recently, there has been a growing interest in fascinating phenomena in $5d$ TMOs, which originate from the competition of spin-orbit coupling (SOC), on-site Coulomb interaction, and crystal-field splitting. Well-established materials in which SOC play crucial roles include the $J_{\text{eff}} = 1/2$ Mott insulator $Sr_2IrO_4$ ($d^5$ electron configuration) [1], the Kitaev quantum spin-liquid compound $Na_2IrO_3$ ($d^5$) [2,3], the all-in/all-out magnet $Cd_2Os_2O_7$ ($d^3$) and $R_2Ir_2O_7$ ($R$ = rare earth) ($d^5$) [4–8], the ferromagnetic Mott-insulator $Ba_2NaOsO_6$ ($d^1$) [9–11], and the spin-orbit coupled metal $Cd_2Re_2O_7$ ($d^2$) [12–14].

The array of aforementioned materials comprises of an octahedral crystal-field on the metal site. In contrast, especially desirable but yet to be performed is an investigation of $5d$ systems with a tetrahedral crystal-field as a counterpart to the octahedral one. The tetrahedral crystal-field yields a doubly-degenerate $e$ level lower and a triply-degenerate $t_2$ level higher in energy, and vice versa in the octahedral crystal field. In metal-oxygen tetrahedra, the metal site loses inversion symmetry locally, from which an antisymmetric spin–orbit coupling (ASOC) emerges as distinct from centrosymmetric metal-oxygen octahedra. Within an $e$ orbital, SOC is usually quenched; however, unlike $t_2$ orbitals it can be activated by $t_2$–$e$ mixing. Hence, we hold that $5d$ TMO with metal-oxygen tetrahedra is a promising platform for studying the interplay among spin, crystal-field, and SOC. Intriguing theoretical works have recently proposed by Song *et al.* [15] and Hayami *et al.* [16] positing that $5d$ TMO scheelite has peculiar magnetic and orbital states governed by crystal-field splitting and SOC, as well as various multipole states arising from the characteristic ASOC; this is detailed in section III.E.

Scheelite is a family of minerals with the chemical formula $ABO_4$ typically found in the tungstate $CaWO_4$ [17,18]. The $A$ site is occupied by a relatively large cation such as an alkali metal or alkali earth metal, while the *B* site is occupied by a high-valence 4*d*- or 5*d*-transition metal (for details see Section III.A.). The *B* atom forms a $BO_4$ tetrahedron, where the $BO_4$ tetrahedra are separated by *A* atoms without sharing common O atoms. The *B* (and also *A*) atoms form a distorted diamond sublattice expanded by 50% along the *c* axis. The global symmetry of this structure is centrosymmetric, while the local spatial inversion symmetry is broken at the *A* or *B* site with –4 symmetry, which should activate ASOC.

We focus on the osmium oxides crystallizing in the scheelite structure: $AOsO_4$ (*A* = K, Rb and Cs) contains $Os^{7+}$ ions with a $5d^1$ electron configuration. $AOsO_4$ has been previously synthesized by reaction in an aqueous solution, however, its structural parameters and magnetic properties remain unknown [19]. We prepared powder samples by solid-state reaction. X-ray diffraction analyses revealed that the K and Rb compounds crystallize in the $CaWO_4$-type scheelite structure, and the Cs compound crystallize in a quasi-scheelite $CsIO_4$ structure [20]. We found antiferromagnetic transitions at $T_N$ = 36.9 and 21.0 K for the K and Rb compounds, respectively, while the Cs compound showed no magnetic ordering above 2 K. Moreover, only the Cs compound exhibits a first-order structural phase transition at $T_s$ = 152.5 K with a tiny change in magnetic interaction.

**II. EXPERIMENTAL**

We prepared powder samples of $AOsO_4$ (*A* = K, Rb, Cs) from $A_2CO_3$ (*A* = K, Rb, Cs), Os, and AgO (as the oxygen source) in a molar ratio of 1:2.2:3.5 in a sealed quartz tube at 673 K for 36 h. After grinding the samples, they were fired in oxygen at 2 atm with AgO to remove impurities, in particular, the $AOs_2O_6$ superconductors, in a sealed quartz tube at 673 K for 36 h. The K compound was gently hygroscopic while the others were stable in air. The as-prepared samples are dark brown in color and insulating in resistivity.

We performed x-ray powder diffraction on a Bragg-Brentano diffractometer (Rigaku RINT) with Cu-*K*α radiation at 300 K. The crystal structures were determined using the Rietveld method on RIETAN-FP software, as shown in Fig. 1 [21]. The crystallographic parameters were refined based on the scheelite structure, as previously mentioned [19], giving $R_{wp}$ = 13.9% and 8.9% for K and Rb compounds, respectively. The structure of Cs compound was determined based on a pseudo-scheelite $CsIO_4$-type structure, giving the final $R_{wp}$ = 7.0%. The crystallographic parameters are listed in Table I. Moreover, we carried out synchrotron x-ray powder diffraction to analyze the Cs compound over the temperature range from 50 K to 300 K on a beamline BL-8A at the Photon Factory, High Energy Accelerator Research Organization (KEK). Two-dimensional images were obtained using a diffractometer with a curved imaging plate (Rigaku R-AXIS) at wavelength of $\lambda$ = 0.689278 Å. The images were integrated to yield $2\theta$–intensity data on a DISPLAY software (Rigaku).

The magnetic susceptibilities were measured over the temperature range from 2 K to 350 K in magnetic fields up to 5 T using a SQUID magnetometer (MPMS, Quantum Design Inc.). The heat capacity data were collected over the temperature range from 2 K to 300 K under magnetic fields of 0 and 9 T using a physical property measurement system (PPMS, Quantum Design Inc.). To assess the lattice contributions in heat capacity for K and Rb compounds, we grew non-magnetic single-crystals of $KReO_4$ and $RbReO_4$ by the slow evaporation method [22].

**III. RESULTS and DISCUSSION**

**A. Crystal structures and magnetic interactions of $AOsO_4$ ($A$ = K, Rb, Cs)**

Figures 2(a) and 2(b) show the crystal structures of $AOsO_4$ for $A$ = K and Cs, respectively. The K and Rb compounds crystallize in the $I4_1/a$ space group in a tetragonal crystal system. The $A$ and Os atoms occupy special positions, while the O atom occupies a general position (Table I) [23]. In contrast, the Cs compound crystallizes in a lower-symmetry structure of the $Pnma$ space group in an orthorhombic crystal system. All of the Cs, Os, and three O atoms occupy positions with variable coordinates (Table I) [23].

Figures 2(c)–2(e) focus on the $OsO_4$ tetrahedra for $A$ = K, Rb, and Cs. The Os–O bond lengths are

1.811(9) Å for $A$ = K, 1.759(8) Å for Rb, 1.736(14), 1.737(9), and 1.793(13) Å [1.755(7) Å on average] for Cs. Therefore, the size of OsO$_4$ tetrahedron decreases in the sequence $A$ = K, Rb, and Cs. The O–Os–O angles are $\alpha$ = 114.1(3)° and $\beta$ = 107.2(3)° for K; $\alpha$ = 116.7(3)° and $\beta$ = 106.0(3)° for Rb; $\alpha$ = 108.7(3)°, $\beta$ = 110.8(3)°, $\gamma$ = 111.6(3)°, and $\sigma$ = 103.5(3)° for Cs. Considering the crystal-field, the tetragonal distortion compressed along the $c$ axis with $D_{2d}$ symmetry for $A$ = K and Rb should remove the degeneracy of the $e$ orbital, resulting in the lower $d_{z2}$ and the higher $d_{x2-y2}$ orbitals. The unpaired electron is therefore present in the $d_{z2}$ orbital as previously discussed in the literature [15]. The further distorted tetrahedron for $A$ = Cs is likely to make more complicated $d$ orbitals.

The Os network forms a distorted diamond lattice with Os–Os distances of 4.24 Å for K and 4.42 Å for Rb (Fig. 2). For the Cs compound, the Os network has two alternating Os–Os bonds: a long (4.70 Å) and a short (4.57 Å) distances. The Os–Os–Os bond angles are 83.5° and 123.8° for K, 82.2° and 124.6° for Rb, and 75.4°, 80.9°, and 122.7° for Cs, which deviate considerably from the bond angle of 109.5° in the regular diamond lattice. Hence, the large $A$ cation expands the Os–Os network and shrinks the OsO$_4$ tetrahedron following the sequence $A$ = K, Rb, and Cs.

The Os magnetic moment interacts not directly but through the superexchange pathway of Os–O…O–Os linked by two O atoms [Figs. 2(c)–2(e)]. Along the Os diamond lattice, there are two equivalent pathways [nearest-neighbor (NN) interaction] via the O–O bonds with 3.07(1) Å for K and 3.27(1) Å for Rb, while two inequivalent pathways with 3.31(1) Å and 3.42(1) Å for Cs. Moreover, there are next-nearest-neighbor (NNN) interactions along the [100] and [010] directions with O–O distances of 3.35(1) Å for K, 3.36(1) Å for Rb, and 3.07(1) Å and 3.21(1) Å for Cs.

**B. Magnetic properties of KOsO$_4$ and RbOsO$_4$**

Figure 3 shows the magnetic susceptibility over the temperature range from 2 K to 330 K under a magnetic field of 1 T for the K and Rb compounds. The Curie–Weiss fittings above 180 K give Weiss temperatures $\theta$ = −66 and −18 K, respectively, and effective magnetic moments $\mu_{\text{eff}}$ = 1.44 and 1.45 $\mu_B$/Os, respectively. Antiferromagnetic interactions are therefore dominant in both compounds. The stronger interaction in the K compound is interpreted from the 4% shorter Os–Os distance to the Rb compound. Their effective magnetic moments are rather small compared with 1.73$\mu_B$ in the simple $d^1$ spin ($S$ = 1/2, $g$ = 2). This result implies that the orbital contribution of −0.3$\mu_B$ has the opposite sign to the Os spin. The susceptibilities of the K and Rb compounds feature humps near $T_{max}$ = 60 and 30 K, and antiferromagnetic orderings at $T_N$ = 37.7 and 20.9 K (Fig. 3 inset), respectively. The transition temperatures are defined at the peaks in the temperature derivatives of the susceptibilities. A hump in susceptibility also appears in the $4d^1$ scheelite KRuO$_4$ [24].

In an antiferromagnet on a diamond lattice, a hump in susceptibility appears along with a short-range-ordering similar to a two-dimensional square-lattice antiferromagnet [25]. We plot the calculated susceptibilities in Fig. 3 based on a high-temperature series expansion on $S$ = 1/2 diamond-lattice Heisenberg antiferromagnet [26] with the magnetic interactions of $J$ = −55 and −15 K for $A$ = K and Rb, respectively, holding the $\mu_{eff}$ values from Curie-Weiss fittings. Though the susceptibilities exhibit the strong reductions from the Curie-Weiss law, there are still large differences between the observed and calculated susceptibilities. Thus, the short-range-ordering alone is insufficient to account for the hump of susceptibility. Beside, in a localized electron system with a small crystal-field splitting, a hump often emerges because the excitation gap competes with the corresponding temperature. We in fact estimate the excitation gaps to be 289–230 K in the present compounds, to be detailed below. Therefore, we propose that the susceptibility hump results from thermal excitations between the ground state and the low-lying first excited state.

Figures 4(a) and 4(b) illustrate the temperature dependence of heat capacity for the K and Rb compounds. These data feature broad humps near 55 and 35 K, and pronounced peaks at $T_N$ = 36.9 and 21.0 K, respectively. The observed peak temperatures agree with the $T_N$ values from susceptibility, which evidences that the antiferromagnetic transitions occur at these temperatures. The transition temperatures remain the same even for magnetic fields of 9 T.

In order to estimate the magnetic contributions to the heat capacity, the lattice contributions are estimated from the heat capacities of the non-magnetic $KReO_4$ and $RbReO_4$. After subtracting them from the total values, we obtain the magnetic heat capacities and the magnetic entropies as shown in Figs. 5(a) ($A$ = K) and 5(b) ($A$ = Rb). The magnetic entropies reaches $S_{mag}$ = 2.7 and 3.6 J mol$^{-1}$ K$^{-1}$ up to $T_N$, and $S_{mag}$ = 14.6 and 15.5 J mol$^{-1}$ K$^{-1}$ up to 300 K for the K and Rb compounds, respectively. The former values correspond to 47% and 62% of $R\ln2$ = 5.76 J mol$^{-1}$ K$^{-1}$ expected for an $S$ = 1/2 magnet. The latter values exceed $R\ln2$ and even $R\ln4$ = 11.53 J mol$^{-1}$ K$^{-1}$ for a four-level system with the two $e$ orbitals and are close to $R\ln6$ = 14.90 J mol$^{-1}$ K$^{-1}$ for a six-level system. However, one has to be careful about an error in estimating the lattice contributions in the high-temperature range that may cause overestimate. As a result, the total magnetic entropies may originate from the lifting of entropy in the $e$ orbital system. We here consider that the broadened humps at 110–120 K for both compounds in the magnetic heat capacity derive from the excitations within the split $e$ orbitals. Rough estimates using a standard Schottky functional form give the excitation gaps to be $\Delta_t$ = 289(6) K for K and $\Delta_t$ = 230(13) K for Rb in the trial temperature range of $T$ > 82 K.

**C. Magnetic properties of CsOsO$_4$**

Figure 6(a) plots the temperature dependence of the magnetic susceptibility and its inverse for the Cs compound. The data follows the Curie–Weiss behavior down to 2 K, and no long-range ordering is observed over the entire temperature range. The absence of a susceptibility hump suggests a small $\Delta_t$ gap for Cs. While, a tiny anomaly appears at $T_s$ = 152.5 K [marked by the arrows in Figs. 6(a) and 6(b)]. The slope in the inverse susceptibility varies at $T_s$, resulting in $\theta$ = 12 K with $\mu_{eff}$ = 0.80 $\mu_B$/Os above 160 K and $\theta$ = 3 K with $\mu_{eff}$ = 0.86 $\mu_B$/Os below 150 K.

The weak ferromagnetic interaction and the smaller effective magnetic moment in the Cs compound indicate a magnetic character distinct from those for K and Rb. This difference seems to have two main causes: the weak NN antiferromagnetic interaction along with the longer Os–Os distance, and the stronger NNN (ferromagnetic) interaction arising from a different orientation of the $OsO_4$ tetrahedron. These structural and magnetic aspects suggest that the NNN interaction competes with the NN interaction in the case of the Cs compound.

The smaller magnetic moment for the Cs compound seems responsible for the stronger SOC. The double-perovskite $5d^1$ magnet $Ba_2NaOsO_6$ with the $Os^{7+}O_6$ octahedral units exhibits a similarly small magnetic moment $\mu_{eff} = 0.6$ $\mu_B$/Os by counteracting the spin and orbital moments due to SOC within the $t_{2g}$ orbital [11]. In contrast, the single-electron configuration within the $e$ orbital for the present Cs compound requires a different mechanism; specifically, the $t_2$–$e$ mixing must be crucial for the emerging SOC (see section III.E) [15]. That is, the smaller magnetic moment for the Cs compound may arise from the narrower $t_2$–$e$ crystal-field splitting compared with those of the K and Rb compounds.

Figure 6(c) represents the heat capacity as a function of temperature for the Cs compound. A pronounced peak appears at $T_s$ = 154.8 K, which coincides with the anomaly in susceptibility. The change in entropy near $T_s$ is estimated to be $\Delta S$ = 0.41 J mol$^{-1}$ K$^{-1}$, by subtracting a phenomenological base line. The large value of $\Delta S$ originates from a structural transition; see next section for details.

## D. Structural transition in CsOsO$_4$

We explored the origin of the anomalies at $T_s$ by using low-temperature x-ray diffraction. Figure 7(a) presents synchrotron x-ray powder patterns at 170 and 130 K. At 170 K, we can fit the data to the orthorhombic *Pnma* structure with $a$ = 5.6811(2), $b$ = 5.9459(2), and $c$ = 14.2936(4) Å. At 130 K, the pattern is quite different, indicating a large structural transition at $T_s$. Employing the Le Beil method without a structural model [21], this pattern can be reproduced with a monoclinic cell of $a$ = 5.7718(5), $b$ = 5.8156(5), $c$ = 14.2350(8) Å, and $\gamma$ = 92.627(6)°; in each pattern fit, we assign the $c$ axis to the orthorhombic or monoclinic unique axis [23]. From the extinction rule for reflections, we infer that, at low temperature, the space group may be either $I2/b$, $Ib$, or $I2$. It is noted that related compound $Ln$NbO$_4$ ($Ln$ = Y, La, Dy) exhibits a first-order ferroelastic phase transition from high-temperature scheelite ($I4_1/a$) to low-temperature fergusonite ($I2/b$) [27,28]. Although a similar transition was expected, it was difficult to fit the low-temperature data assuming the fergusonite structure.

Figure 7(b) plots the lattice constants as a function of temperature for the Cs compound. The lattice constants at $T_s$ reveal sudden changes, suggesting a first-order transition. The lattice distortions in the orthorhombic and monoclinic crystal systems are estimated to be $a_o/b_o - 1 = -4.5 \times 10^{-2}$ and $a_m/b_m - 1 = -7.5 \times 10^{-3}$. Thereby, the low-temperature phase has minor lattice distortions in the $ab$ plane. Moreover, the monoclinic angle of $\gamma \sim 92°$ indicates a small shear deformation of the $ab$ plane. The unit cell volume $V$ of the low-temperature phase is 1% less than that of the high-temperature phase, $\Delta V = -4.44$ Å$^3$. Thus, the above findings demonstrate the distinct structural change with the small lattice deformation at $T_s$.

**E. Spin-orbit coupling in $A$OsO$_4$**

The theoretical calculation by Song *et al.* concerning KOsO$_4$ has revealed the electronic and magnetic states as follows [15]. They have presented the $t_2$–$e$ splitting $\Delta_c = 1.7$ eV, the large SOC of 0.3 eV, and the large orbital moment of $-0.2$ $\mu_B$/Os. Generally, SOC within $e$ orbitals is quenched completely. However, SOC can be activated via the $t_2$–$e$ mixing because the crystal-field splitting is small in the tetrahedral configuration of the O atoms (cf. $\Delta_c = 6$ eV for Ba$_2$NaOsO$_6$ with OsO$_6$ octahedron [11]). This is evidenced by the observed large orbital moment of KOsO$_4$. In consequence, the reduction of the effective magnetic moment observed in the susceptibility derives from competing spin and orbital moments. Moreover, the $e$-level splitting above the ground state $d_{z^2}$ level has been calculated to be $\Delta_t = 84$ meV. This value is larger than the excitation gap of 289 K obtained from heat capacity. This feature is likely to remain in the Rb compound having similar properties. In contrast, what are observed for the Cs compound is quite different: much smaller effective moment of 0.86 $\mu_B$/Os below $T_s$, compared with ~1.45 $\mu_B$/Os for the others, and the negligible $e$ level splitting. This must be related to the structural transition only observed for the Cs compound. Thus, an instability toward a multipolar order, as theoretically suggested [16], may drive the $T_s$ transition.

Hayami *et al*. have shown that the staggered spin and orbital order in $A$OsO$_4$ ($A$ = K, Rb, Cs) involves odd-parity multipoles—a magnetic monopole, quadrupoles, and toroidal dipoles [16]. Because the magnitude of ASOC in this system stems from the degree of $t_2$–$e$ mixing, the crystal-field splitting is regarded as a control parameter for the multipole ordering. This implies that chemical or physical pressure can control the magnitude of ASOC. Moreover, they have given detailed classifications of odd-parity multipoles and magnetoelectric effects in the magnetic ordered states [16]. To improve our understanding of this system, we will examine a determination of the magnetic ordered states for the K and Rb compounds and an investigation of their magnetoelectric effects after synthesizing single crystals.

## IV. SUMMARY

We prepared polycrystalline samples of the $5d^1$ transition metal oxides $A$OsO$_4$ ($A$ = K, Rb, Cs) by solid-state reaction, determined their structure, and measured their magnetic properties and heat capacity. The K and Rb compounds form a scheelite structure made of a distorted diamond lattice of Os atoms with OsO$_4$ tetrahedral units. The Cs compound has a quasi-scheelite structure with lower symmetry. For $A$ = K and Rb, the effective magnetic moments are $\mu_{\text{eff}}$ = 1.44 and 1.45 $\mu_B$/Os, respectively, while it is 0.8 $\mu_B$/Os for $A$ = Cs. These reduced values are tied to the orbital moments modified by crystal-field splitting. Antiferromagnetic transitions appear at $T_N$ = 36.9 K for $A$ = K and at 21.0 K for $A$ = Rb, whereas no magnetic transition is observed down to 2 K for the Cs compound. Moreover, we find a structural transition at $T_s$ = 152.5 K for $A$ = Cs. A $5d$ transition metal with a strong SOC on the diamond lattice is a good platform for studying ASOC with locally broken inversion symmetry. With their single-crystal synthesis, $A$OsO$_4$ compounds will be good candidates for revealing new aspects of ASOC related to the synergy between the crystal field and the orbital degrees of freedom as well as the magnetoelectric effects.

**ACKNOWLEDGEMENTS**

We thank S. Hayami, Y. Motome, K. Ohgushi, H. Harima, and Y. Kuramoto for fruitful discussions. This work was supported by MEXT Element Strategy Initiative to Form Core Research Center and JSPS KAKENHI (No. 16K05434). This work was performed using facilities of the Institute for Solid State Physics, the University of Tokyo. The synchrotron x-ray study was performed with approval of the Photon Factory Program Advisory Committee (No. 2013S2-002 and No. 2016S2-004). All the crystal structures are drawn by using the software VESTA [29].

TABLE I. Crystallographic data for $A$OsO$_4$ at 300 K. The structural parameters are refined with the $I4_1/a$ (#88) space group for KOsO$_4$ and RbOsO$_4$, and $Pnma$ (#62) for CsOsO$_4$. The atomic positions and Wyckoff letters (W. L.) are also given [23].

| Aom | W. L. | $x$ | $y$ | $z$ |
|---|---|---|---|---|
| K; $a$ = 5.65206(5) Å, $c$ = 12.6642(2) Å | | | | |
| K | 4$b$ | 0 | 1/4 | 5/8 |
| Os | 4$a$ | 0 | 1/4 | 1/8 |
| O | 16$f$ | 0.1318(14) | 0.0156(15) | 0.2028(7) |
| Rb; $a$ = 5.82031(8) Å, $c$ = 13.3324(3) Å | | | | |
| Rb | 4$b$ | 0 | 1/4 | 5/8 |
| Os | 4$a$ | 0 | 1/4 | 1/8 |
| O | 16$f$ | 0.1092(13) | 0.0170(13) | 0.1943(6) |
| Cs; $a$ = 5.75256(8) Å, $b$ = 5.93881(8) Å, $c$ = 14.3458(2) Å | | | | |
| Cs | 4$c$ | 0.0245(2) | 3/4 | 0.1266(2) |
| Os | 4$c$ | 0.0373(2) | 1/4 | 0.3796(1) |
| O1 | 4$c$ | 0.8184(24) | 1/4 | 0.07635(9) |
| O2 | 4$c$ | 0.8370(21) | 1/4 | 0.4754(9) |
| O3 | 8$d$ | 0.0113(15) | 0.5203(14) | 0.6928(6) |

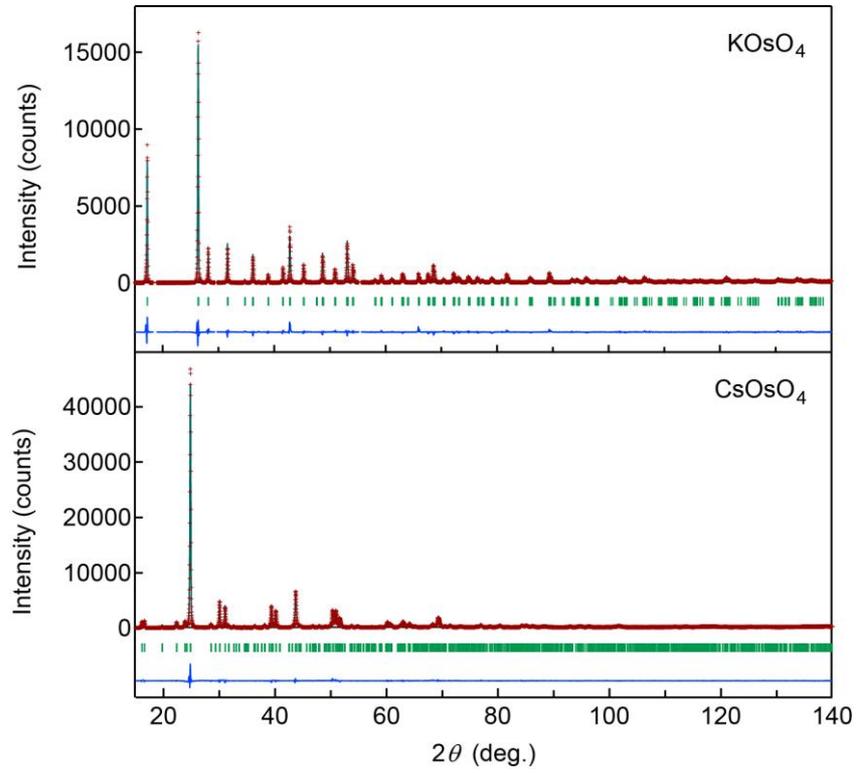

FIG. 1. Powder x-ray diffraction patterns of KOsO$_4$ (upper) and CsOsO$_4$ (lower) with Cu-$K\alpha$ radiation at 300 K. The plots correspond to the observed (red crosses), calculated (black line), difference profiles (blue line), and positions of Bragg peaks (green tick marks).

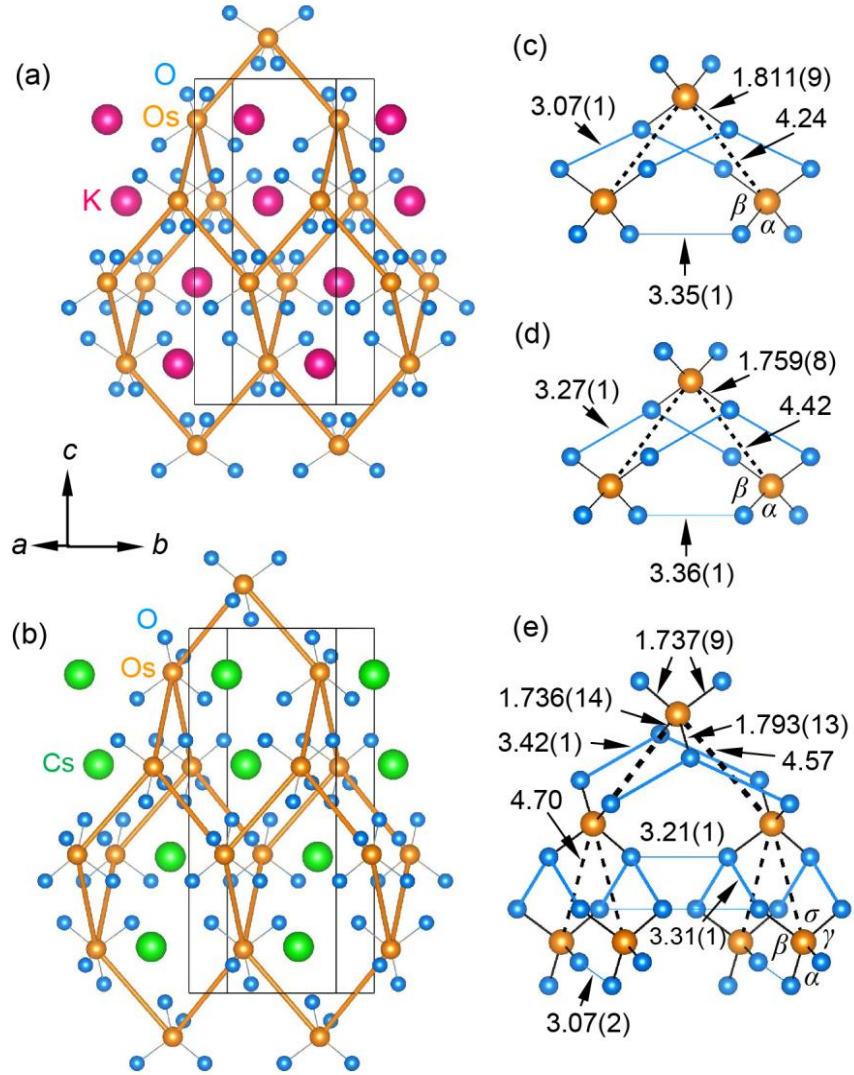

FIG. 2. Crystal structures of KOsO$_4$ (a) and CsOsO$_4$ (b). Each Os atom is surrounded tetrahedrally by four oxide atoms. The distorted diamond networks made of Os atoms are drawn with sticks. The Os–Os, Os–O, and O–O distances in Angstroms are written for KOsO$_4$ (c), RbOsO$_4$ (d), and CsOsO$_4$ (e). The diamond lattices are represented with broken lines for K and Rb, while for CsOsO$_4$, the short and long Os–Os distances are represented by thick and thin broken lines, respectively. The $\alpha$–$\sigma$ listed on the OsO$_4$ tetrahedron are the O–Os–O bond angles, whose values are given in the text.

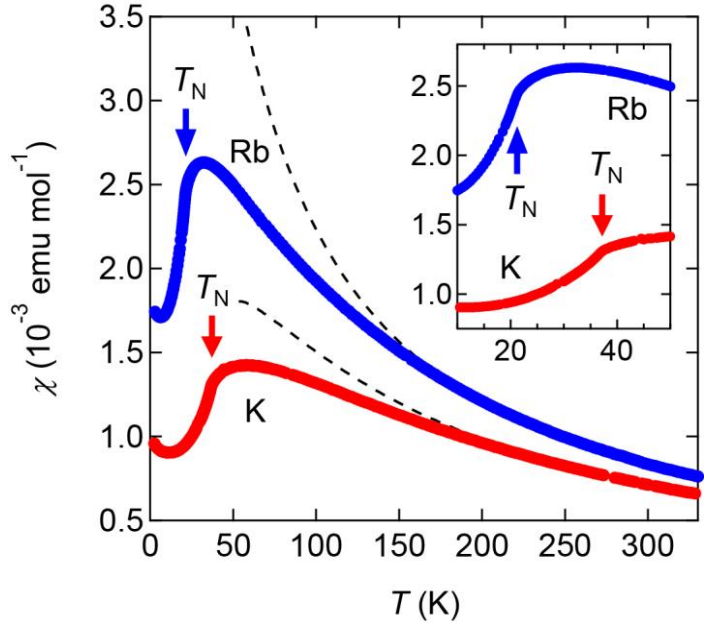

FIG. 3. Temperature dependence of magnetic susceptibility down to 2 K obtained under a magnetic field of 1 T for $KOsO_4$ (red) and $RbOsO_4$ (blue). A molar unit corresponds to one formula unit of $AOsO_4$. Broad humps appear near 60 and 30 K for $A$ = K and Rb, followed by antiferromagnetic orders at $T_N$ = 37.2 and 21.2 K, respectively (indicated by arrows). The broken lines represent the calculated susceptibilities based on a high-temperature series expansion on $S = 1/2$ diamond lattice Heisenberg antiferromagnetic model [26]. The inset shows enlargements of magnetic susceptibility for $A$ = K and Rb around $T_N$.

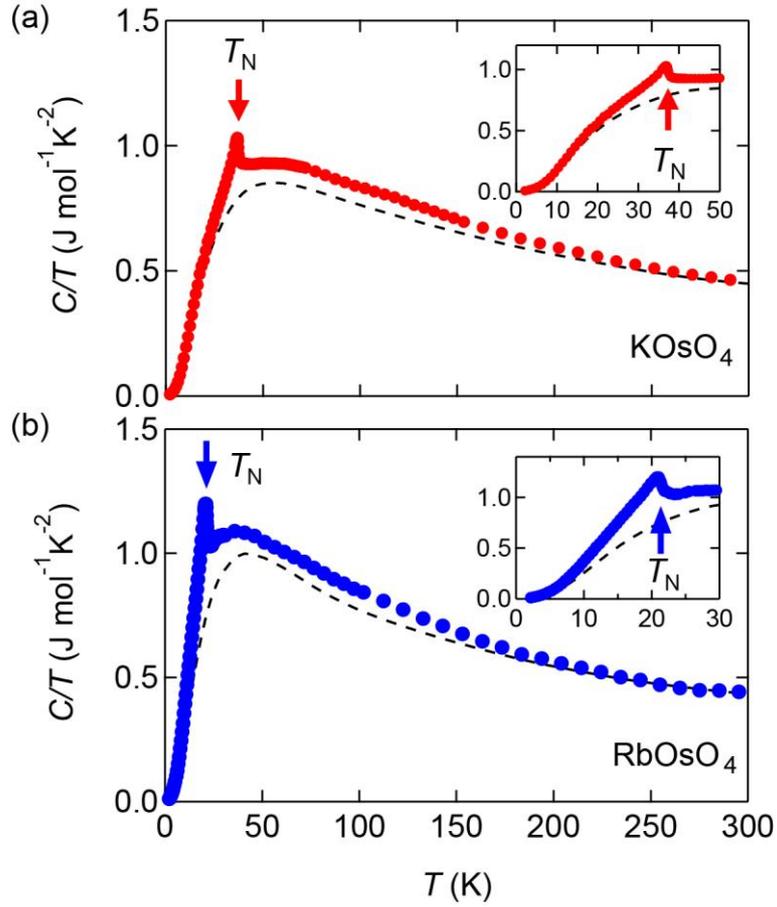

FIG. 4. Temperature dependence of heat capacity divided by temperature, $C/T$, for $KOsO_4$ (a) and $RbOsO_4$ (b). A molar unit corresponds to one formula unit of $AOsO_4$. As references of lattice contributions, the $C/T$ of nonmagnetic $KReO_4$ and $RbReO_4$ are plotted with broken lines in (a) and (b), respectively. Sharp peaks indicative of antiferromagnetic ordering appear at $T_N$ = 36.9 and 21.0 K for $KOsO_4$ and $RbOsO_4$, respectively.

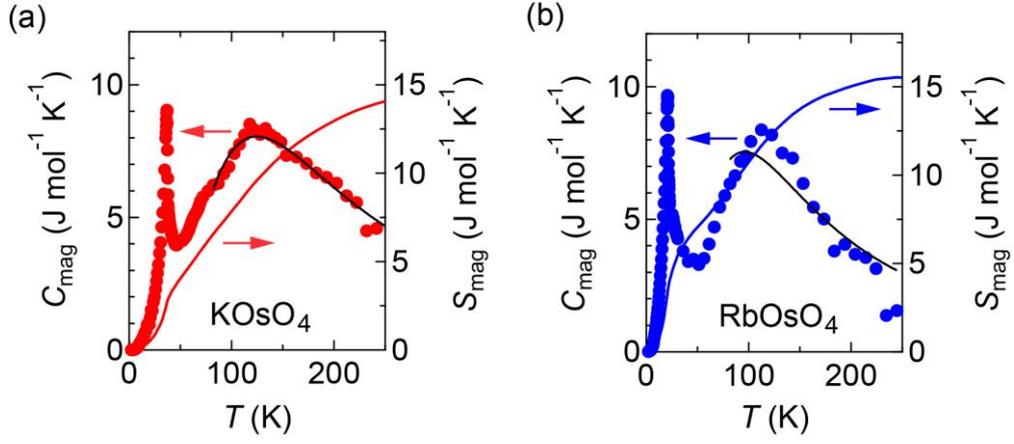

FIG. 5. Temperature dependence of magnetic heat capacity, $C_{\text{mag}}$, and magnetic entropy, $S_{\text{mag}}$, for $KOsO_4$ (a) and $RbOsO_4$ (b). A molar unit corresponds to one formula unit of $AOsO_4$. Assuming the humps to be of Schottky type in $C_{\text{mag}}$, the excitation gaps are estimated to be $\Delta_t = 289(6)$ K for K and $\Delta_t = 230(13)$ K for Rb, as drawn by the black solid lines.

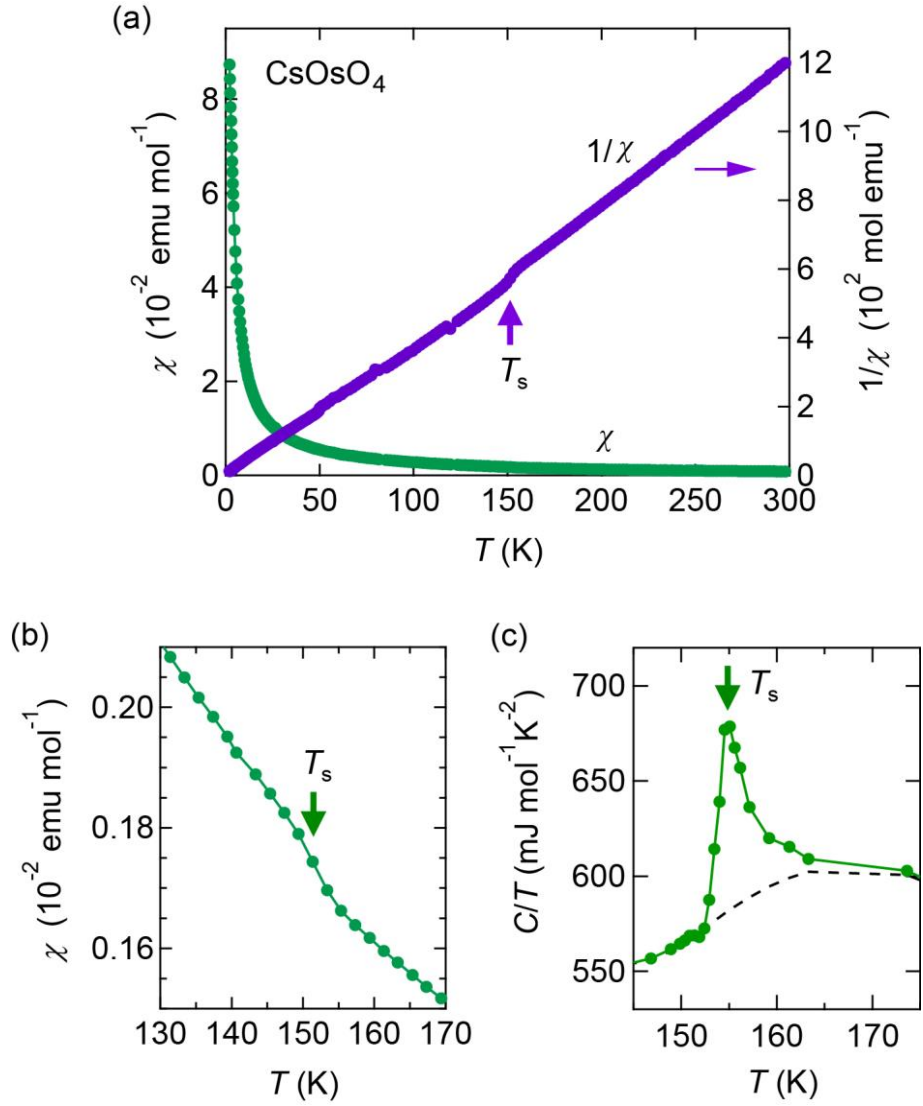

FIG. 6. (a) Temperature dependence of magnetic susceptibility and inverse magnetic susceptibility measured at 1 T for $CsOsO_4$. The molar unit corresponds to one formula unit of $CsOsO_4$. A weak anomaly appears at $T_s$ = 152.5 K, where the inverse susceptibility changes in gradient. (b) Magnetic susceptibility near $T_s$. (c) Heat capacity divided by temperature, $C/T$, for $CsOsO_4$, featuring a jump at $T_s$ = 154.8 K with a change in entropy of $\Delta S$ = 0.41 J mol$^{-1}$ K$^{-1}$, determined from the approximate base line (broken line).

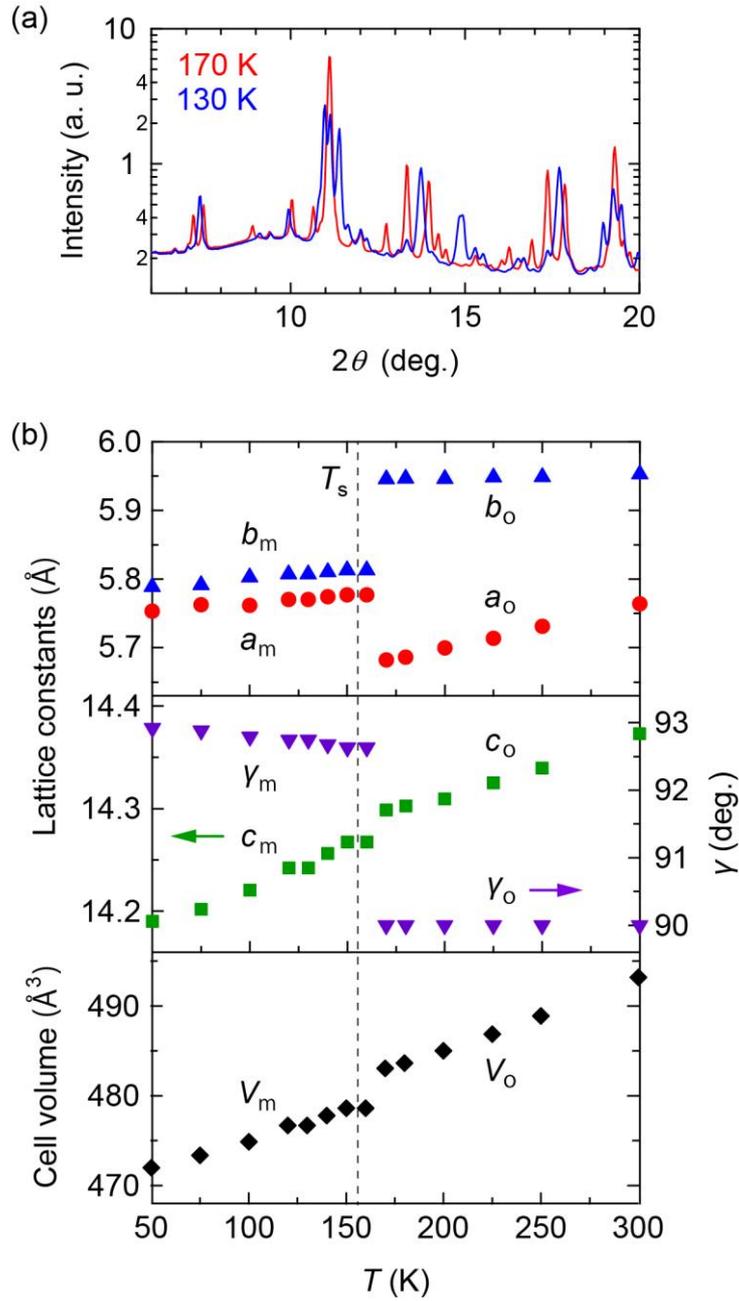

FIG. 7. (a) Synchrotron powder x-ray diffraction pattern for $CsOsO_4$ at 170 K (red solid line) and 130 K (blue solid line) at wavelength of $\lambda = 0.689278$ Å. (b) A large structural change at $T_s = 154.8$ K is seen in temperature dependence of the lattice constants $a$ (red circle), $b$ (blue upward triangle), $c$ (green square), and $\gamma$ (purple downward triangle), and $V$ (black diamond). Subscripts o and m denote orthorhombic and monoclinic crystal systems, respectively.